\documentclass[%
reprint,
superscriptaddress,
 amsmath,amssymb,
 aps,
]{revtex4-2}

\usepackage{graphicx}
\usepackage{dcolumn}
\usepackage{bm}
\usepackage{physics}
\usepackage{booktabs}
\usepackage{subfig}
\usepackage[ruled]{algorithm2e}
\usepackage{multirow}

\usepackage[normalem]{ulem} 
\usepackage{amsmath}
\usepackage{xcolor}         
\usepackage{soul}           
\usepackage{changes}        

\definecolor{added}{RGB}{0,100,0}      
\definecolor{deleted}{RGB}{150,0,0}    
\definecolor{comment}{RGB}{0,0,150}    

\preprint{APS/123-QED}

\begin{document}

\title{Improving neural network performance for solving quantum sign structure}

\author{Xiaowei Ou}
 \email{xiaowei.ou@yale.edu}
\affiliation{Department of Physics, Yale University, New Haven, Connecticut 06520, USA}
\affiliation{Energy Sciences Institute, Yale University, West Haven, Connecticut 06516, USA}
\author{Tianshu Huang}
\affiliation{Energy Sciences Institute, Yale University, West Haven, Connecticut 06516, USA}
\affiliation{Department of Applied Physics, Yale University, New Haven, Connecticut 06520, USA}
\author{Vidvuds Ozoli\c{n}\v{s}}%
 \email{vidvuds.ozolins@yale.edu}
\affiliation{Energy Sciences Institute, Yale University, West Haven, Connecticut 06516, USA}
\affiliation{Department of Applied Physics, Yale University, New Haven, Connecticut 06520, USA}

\date{\today}

\begin{abstract}
Neural quantum states have emerged as a widely used approach to the numerical study of the ground states of non-stoquastic Hamiltonians. However, existing approaches often rely on {\it a priori\/} knowledge of the sign structure or require a separately pre-trained phase network. We introduce a modified stochastic reconfiguration method that effectively uses differing imaginary time steps to evolve the amplitude and phase. Using a larger time step for phase optimization, this method enables a simultaneous and efficient training of phase and amplitude neural networks. The efficacy of our method is demonstrated on the Heisenberg $J_1-J_2$ model.

\end{abstract}

\maketitle


\section{Introduction}
Finding ground states in strongly interacting many-body systems is an important problem due to its connection to many intriguing phenomena in condensed matter physics, such as quantum spin liquids \cite{zhou2017quantum,savary2016quantum}, high $T_c$ superconductors \cite{phillips2012physics}, and topological insulators \cite{ando2013topological}. However, analytical solutions are rarely possible \cite{Levkovich-Maslyuk_2016}, and the exponential growth of the Hilbert space with respect to the system size presents challenges for numerical studies.

Many numerical methods have been developed to address specific classes of problems efficiently, but with characteristic limitations. For instance, exact diagonalization methods \cite{PhysRevB.39.4744,PhysRevB.79.195102} are applicable only to small system sizes. Quantum Monte Carlo  \cite{doi:10.1126/science.231.4738.555,becca2017quantum} encounters difficulties in frustrated and fermionic systems due to the infamous sign problem \cite{troyer2005computational}. The density matrix renormalization group \cite{schollwock2005density} is very successful in low-dimensional systems, but it becomes cumbersome in higher dimensions. Complementary to all these are variational approaches, which aim to efficiently parametrize the physically relevant part of the Hilbert space. Various methods lie in this category, such as Jastrow- and Gutzwiller-projected wave functions \cite{tahara2008variational,PhysRevB.97.235103}, tensor network states \cite{PhysRevB.94.075143,orus2019tensor}, projected entangled pair states \cite{PhysRevB.96.205152,liang2021hybrid} and, most recently, neural quantum states (NQS).

Neural quantum states were first introduced using restricted Boltzmann machine as a practical wave-function Ansatz \cite{carleo2017solving}. Since then, this neural network architecture has been widely studied \cite{borin2020approximating,gao2017efficient,Viteritti_2022,carleo2018constructing}. Apart from the restricted Boltzmann machine, a variety of other architectures have been explored, including fully connected neural networks \cite{saito2018machine}, convolutional neural networks \cite{saito2018machine,choo2019two,yang2020deep,PhysRevB.108.054410,liang2018solving}, recurrent neural networks \cite{hibat2020recurrent}, autoregressive models \cite{sharir2020deep}, transformer neural networks \cite{viteritti2023transformer}, and various neural network inspired Ans\"azte with physical information \cite{ferrari2019neural,PhysRevResearch.4.L012010,nomura2017restricted,robledo2022fermionic,luo2019backflow,inui2021determinant}. These architectures have been used to study ground-state and the lowest excited-state \cite{choo2018symmetries} properties in strongly correlated one- (1D) and two-dimensional (2D) systems.

Neural quantum states have several advantages over other variational Ans\"azte. First, because of the universal approximation theorems, a neural network with a sufficient number of parameters can represent any function \cite{ZHOU2020787}, hence in principle it does not have any variational bias. Second, neural quantum states can be easily generalized to higher-dimensional systems and larger system sizes \cite{PhysRevB.108.054410}. Third, NQS requires sub-exponential number of parameters to encode volume-law entangled states \cite{deng2017quantum,sehayek2019learnability,Viteritti_2022,sun2022entanglement}.

In spite of all these advantages, many challenges remain when using neural quantum states. For instance, representing the non-trivial sign structure in frustrated and fermionic systems is still a difficult problem \cite{PhysRevB.107.195115,PhysRevB.97.035116,PhysRevResearch.4.L022026,PhysRevResearch.3.043126,PhysRevResearch.2.033075,westerhout2020generalization}. Furthermore, due to a rugged landscape in the parameter space \cite{10.21468/SciPostPhys.10.6.147}, optimization of neural network parameters is often difficult and requires substantial computational resources.

In this work, to alleviate the problem of optimizing non-trivial sign structure, we propose an improved stochastic reconfiguration method (SR), which introduces different learning rates for the optimization of the amplitude and phase of the wave function. By employing a faster learning rate for the phase, our proposed method effectively balances the contributions from the phase and amplitude to the gradients in the network parameter space. This algorithm is tested on the paradigmatic Heisenberg $J_1-J_2$ model on a $6 \times 6$ square lattice, and they are shown to perform reliably and efficiently.

\section{Neural quantum states}
In a spin lattice system with N sites, a quantum state can be expanded as $\ket{\psi}=\sum_\sigma \psi_\theta(\sigma) \ket{\sigma}$, where $\sigma = (\sigma_1, \sigma_2, ..., \sigma_N)$ are $\{\hat{S}_i^z\}$ eigenstates spanning the Hilbert space of the spin configurations, and $\theta$ denotes the vector of variational parameters. In neural quantum states, the Ansatz function $\psi_\theta (\sigma)$ is given by an artificial neural network, where the input is a basis spin configuration $\sigma$, encoded by a 2D matrix with values $+1$ $(-1)$ denoting up (down) spin states. It is common practice to have the neural network output the logarithm of the wave function, so that it can accurately capture coefficients over multiple orders of magnitude. In this work we use
\begin{equation}
    \ln [\psi_\theta (\sigma)] = A_\theta(\sigma) + i \phi_\theta(\sigma),
\end{equation}
where $A_\theta(\sigma)$ and $\phi_\theta(\sigma)$ are outputs from two independent real-valued neural networks. It has been noted that the most accurate results are usually obtained with networks that do not make such a separation but allow the flow of information between the phase and amplitude.  Our proposed methods in Sections \ref{sec:time-step-adjusted-optimization} remain applicable to these more complex architectures and they can support both real-valued and complex-valued network parameters.

\begin{figure}[htbp]
    \centering
    \includegraphics[width=0.48 \textwidth]{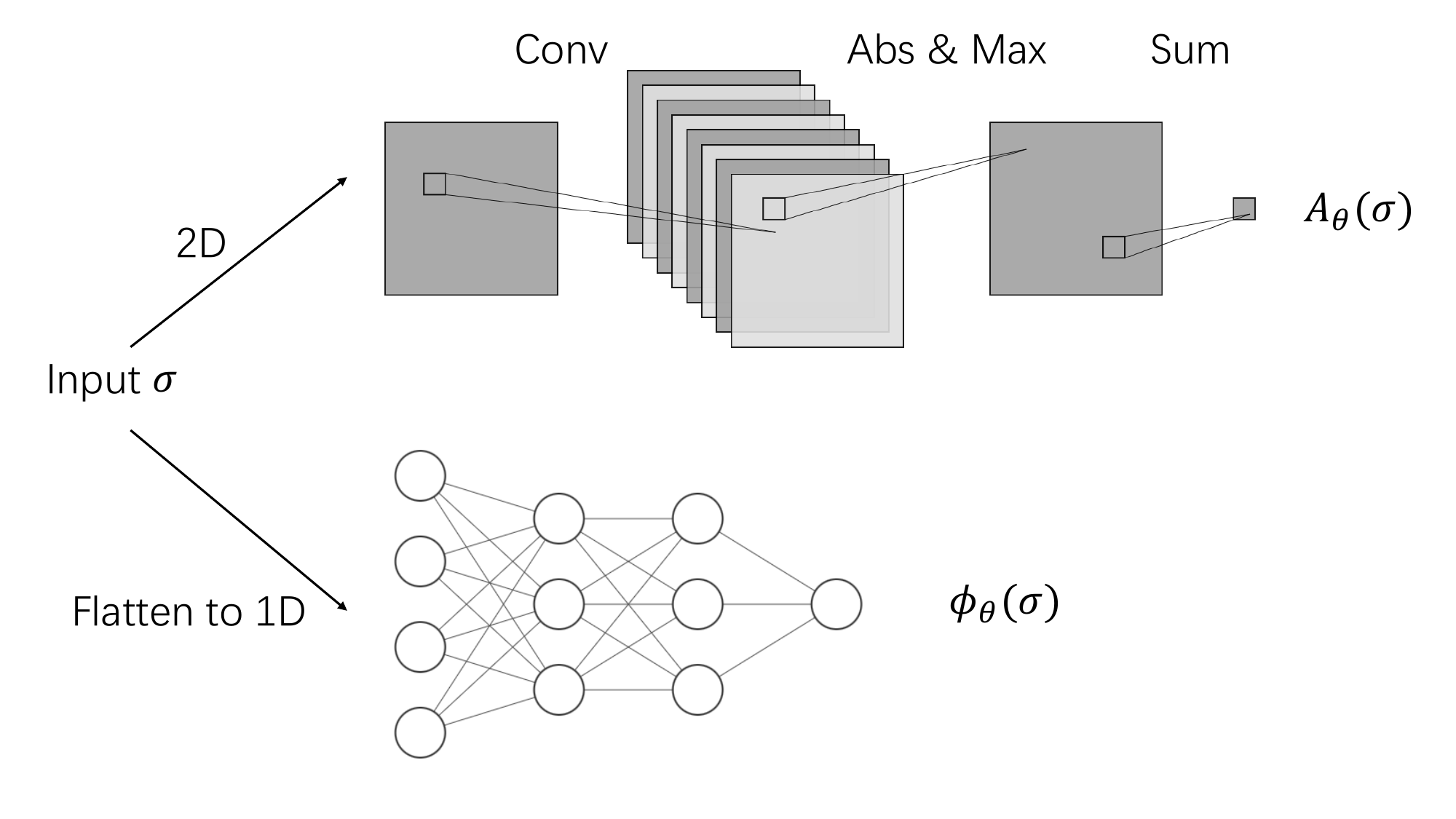}
    \caption{Neural network architecture.}
    \label{fig:neural_network}
\end{figure}

Figure \ref{fig:neural_network} shows the neural network architecture. For $A_\theta(\sigma)$, the 2D configuration first passes through a convolutional layer with 64 $6 \times 6$ kernels. We then take absolute value for each element, followed by taking the maximum value for each site across the 64 kernels, and then sum over all sites to get the output. This can be formally written as
\begin{equation}
    A_\theta(\sigma) = \sum_{ij}\max_{c}\left |\sum_{lm}w_{l,m}^c \sigma_{i+l,j+m}+b^c\right |,
\end{equation}
where $(i,j)$ index lattice sites in the input configuration $\sigma$, $c$ is the channel number, and $(l,m)$ index kernel sites after the convolution. Tensors $w^c$ are the kernel weights, and $b^c$ are biases.

For the phase $\phi_\theta(\sigma)$ we use three fully-connected layers each followed by ReLU ($\text{ReLU}(x) = \max(0,x)$), except after the last layer. The number of neurons in the hidden layers are both 8, and the number of neurons in the output layer is 1. In total, our neural networks have 2745 real-valued parameters. We use 2000 Monte Carlo samples and optimize the network using Nesterov's accelerated gradient method \cite{10.5555/3042817.3043064} with momentum coefficient $\mu=0.5$ and learning rate of $0.04$. Our work uses a simple architecture with reasonably good but not the best possible performance in order to focus on the method for optimizing neural quantum states.

We further enforce the symmetry of the $C_{4v}$ point group. Since the ground state is fully symmetric with a total wave vector $\mathbf{k}=\mathbf{0}$ \cite{nomura2021helping}, we use the characters of the $A_1$ irreducible representation. In the case where the group representation $A_1$ is one-dimensional, the symmetrized wave function can be written as 

\begin{equation}
\label{eq:symm}
    \psi_\theta(\sigma) = \frac{1}{|G|} \sum_{g \in G} \chi(g) \psi_\theta(\hat{g} \sigma)
\end{equation}
where $g$ is a group element, $\chi(g)$ is the character corresponding to $g$, and $G$ is the point group $C_{4v}$. Translational symmetry is automatically enforced for the amplitude by the convolutional layer in the net.

We have also tested an approach that uses explicitly symmetrized basis functions instead of $\ket{\sigma}$. This method may prove particularly useful when one is interested in obtaining states of given symmetry at $\mathbf{k} \neq \mathbf{0}$. It is described in Appendix \ref{sec:symmetrized_wave_function}, and its performance is found to be very similar to Eq.~(\ref{eq:symm}).

\section{Stochastic Reconfiguration}
The Wick-rotated Schrödinger equation governs imaginary time evolution according to
\begin{equation}
    \frac{\partial \ket{\psi(\tau)}}{\partial \tau}=-(\mathcal{H}-E)\ket{\psi(\tau)},
    \label{eq:imagtimeSE}
\end{equation}
where $E =\bra{\psi(\tau)}\mathcal{H}\ket{\psi(\tau)}$ is the energy expectation value, introduced to preserve the normalization constraint $\braket{\psi}{\psi}=1$. Numerically, $E$ is estimated as a Monte Carlo average over $N_s$ samples drawn from the probability distribution $|\psi_\theta(\sigma)|^2$: $E = \langle E_{\text{loc}}(\sigma) \rangle$. Here, $E_{\text{loc}}(\sigma) = \sum_{\sigma'} \mathcal{H}_{\sigma \sigma'} \frac{\psi_\theta(\sigma')}{\psi_\theta(\sigma)}$ represents the local energy, and the wave function $\psi_\theta(\sigma)$ is parameterized by $N_p$ variational parameters $\theta$.

The imaginary time evolution induces an infinitesimal change in the wave function components 
\begin{equation}
    \delta \psi_\sigma = -\psi_\sigma \left( E_{\text{loc}}(\sigma) - E \right) \delta \tau.
\end{equation}
On the other hand, parameter variations generate changes through $\sum_k \frac{\partial \psi_\sigma}{\partial \theta_k} \delta \theta_k$. Since we require this change to lie within the tangent space of $\ket{\psi}$, we apply the projection operator $\hat{P}=\mathbb{I} - \ket{\psi}\bra{\psi}$, yielding
\begin{align}
    \delta \psi'_\sigma =\psi_\sigma \sum_k (O_{\sigma k} - \langle O_{\sigma k} \rangle) \delta\theta_k,
\end{align}
where $O_{\sigma k}=\frac{\partial \ln \psi_\theta(\sigma)}{\partial \theta_k}$ is the logarithmic derivative of the wave function. 

To determine the optimal parameter updates $\delta \theta_k$, we minimize the distance between the imaginary-time-evolved state change $\delta \psi$ and the variational state change $\delta \psi'$
\begin{equation}
\begin{aligned}
    d^2 = \lVert \delta \psi' - \delta \psi \rVert^2 &= \lVert \bar{O} \delta \theta - \bar{\epsilon} \rVert^2,
\end{aligned}
\end{equation}
where we have defined the $N_s$-dimensional vector $\bar{\epsilon}_\sigma=-\delta \tau (E_{\text{loc}}(\sigma) - E )/\sqrt{N_s}$, and the $N_s \times N_p$ matrix $\bar{O}_{\sigma k}=(O_{\sigma k}- \langle O_{\sigma k} \rangle)/ \sqrt{N_s}$.

Since the matrix $\bar{O}$ is typically ill-conditioned or singular, Tikhonov regularization is employed to stabilize the optimization, which leads to the following Lagrangian:
\begin{equation}
    \mathcal{L}_{\text{SR}}=\beta \lVert \delta \theta \rVert^2 + \lVert \bar{O}\delta \theta - \bar{\epsilon} \rVert^2.
\end{equation}
Setting $\partial \mathcal{L}_{SR} / \partial(\delta \theta^\dagger)=0$ gives the parameter update
\begin{equation}
    \delta \theta = (\bar{O}^\dagger \bar{O} + \beta \mathbb{I})^{-1} \bar{O}^\dagger \bar{\epsilon}.
\end{equation}
This formula, which we refer to as stochastic reconfiguration, requires inverting an $N_p \times N_p$ matrix. For most modern NQS architectures with a large number of parameters one has $N_s \ll N_p$, and the Woodbury matrix identity \cite{rende2024simple} provides a computationally more efficient alternative:
\begin{equation}
    \delta \theta = \bar{O}^\dagger(\bar{O}\bar{O}^\dagger + \beta \mathbb{I})^{-1}\bar{\epsilon}.
    \label{eq:minsr-regularized}
\end{equation}
This so-called MinSR formula was first proposed by
Chen and Heyl \cite{chen2024empowering} and  only requires inverting a much smaller $N_s \times N_s$ matrix. Chen and Heyl derived the MinSR method using the following Lagrangian:
\begin{equation}
    \mathcal{L}_{\text{MinSR}} = \lVert \delta \theta \rVert^2 + (\delta \theta^\dagger \bar{O}^\dagger - \bar{\epsilon}^\dagger)\mu,
\end{equation}
where $\mu$ is the Lagrangian multiplier. The solution is $\delta \theta = \bar{O}^\dagger (\bar{O} \bar{O}^\dagger)^{-1} \bar{\epsilon}$. In Ref.~\cite{chen2024empowering} regularization is introduced as a numerical stabilization step to arrive at Eq.~(\ref{eq:minsr-regularized}). We note that in real computation, the constraint $\bar{O} \delta \theta = \bar{\epsilon}$ cannot always be satisfied due to the approximate nature of the NQS, especially when approaching the true ground state. To account for this, we incorporate a residue term $\Delta$ into the constraint, leading to the modified Lagrangian:
\begin{equation}
    \mathcal{L}_{\text{MinSR}}' = \beta \lVert \delta \theta \rVert^2 + (\delta \theta^\dagger \bar{O}^\dagger - \bar{\epsilon}^\dagger - \Delta^\dagger)\mu + \lVert \Delta \rVert^2.
\end{equation}
By minimizing this Lagrangian with respect to $\delta \theta$ and $\Delta$ (see the Supplementary Material), we naturally recover the regularized solution given in Eq.~(\ref{eq:minsr-regularized}), providing an alternative justification for Tikhonov regularization in the MinSR framework.

\section{Time-step adjusted optimization}
\label{sec:time-step-adjusted-optimization}

The stochastic reconfiguration method faces challenges when training log-amplitude and phase networks simultaneously, particularly for systems with unknown sign structure, such as frustrated and fermionic systems. The faster changes of log-amplitude affect the Markov chain greatly, impeding the phase network's ability to converge to a correct sign structure.

To illustrate the issue, consider a minimal example with two basis states, $\ket{\psi} = (\psi_1, \psi_2)^T$, where $\psi_1$ and $\psi_2$ are real-valued for the ground state. The energy expectation value is
\begin{equation}
    E = h_{11}\psi_1^2 + h_{22}\psi_2^2 + (h_{12}+h_{21}) \psi_1 \psi_2.
\end{equation}
In the Heisenberg model, off-diagonal elements $h_{12}=h_{21} \geq 0$ require $\psi_1 \psi_2 \leq 0$ for energy minimization. However, if the phase network fails to predict the correct signs (yielding $\psi_1 \psi_2 \geq 0$), the log-amplitude network aggressively suppresses either $|\psi_1|$ or $|\psi_2|$ to eliminate the positive cross term. This biases the Markov chain toward configurations with the smallest diagonal energy $h_{ii}$, preventing exploration of the true ground state. Existing approaches for the two-dimensional Heisenberg $J_1-J_2$ model prioritize phase training \cite{PhysRevResearch.2.033075} or incorporate specific rules into the Ansatz \cite{choo2019two,PhysRevB.107.195115} to address this issue. 

To balance the training speeds of log-amplitude and phase networks, we can decouple the imaginary time evolution for amplitude and phase by advancing them at different rates. Using $\psi_\sigma = A_\sigma e^{i\phi_\sigma}$, we can rewrite Eq.~\eqref{eq:imagtimeSE} as
\begin{align}
    \frac{1}{A_\sigma}\frac{\partial A_\sigma}{\partial \tau} &= -\Re \left[ E_{\text{loc}}(\sigma)-E\right] \label{eq:imaginary-time-evolution-amplitude}, \\
    \frac{\partial \phi_\sigma}{\partial \tau} &= -\Im \left[ E_{\text{loc}}(\sigma)-E \right]. \label{eq:imaginary-time-evolution-phase}
\end{align}
The amplitude step $\delta \tau_A$ is limited by the nonlinearity implicit in the Monte Carlo average on the r.h.s. of Eq.~(\ref{eq:imaginary-time-evolution-amplitude}) which requires one to resample the Markov chain upon large changes in $A_\sigma$. In contrast, the phase time step $\delta \tau_\phi$ can be larger, as advancing the phase using Eq.~(\ref{eq:imaginary-time-evolution-phase}) does not affect the Monte Carlo sampling distribution,
\begin{align}
    A_\sigma^{(n+1)} &= A_\sigma^{(n)}+\delta \tau_A \frac{\partial A_\sigma}{\partial \tau}, \\
    \phi_\sigma^{(n+1)} &= \phi_\sigma^{(n)}+\delta \tau_\phi \frac{\partial \phi_\sigma}{\partial \tau},
\end{align}
where $\delta \tau_A \leq \delta \tau_\phi$. This leads to the change in the wave function:

\begin{align}
    \delta \psi_\sigma &= \psi_\sigma \left(\frac{1}{A_\sigma}\frac{\partial A_\sigma}{\partial \tau} \delta \tau_A + i\frac{\partial \phi_\sigma}{\partial \tau} \delta \tau_\phi \right).
\end{align}

As detailed in the Supplementary Material, the distance between $\delta \psi'$ and $\delta \psi$ is given by
\begin{equation}
    d^2 = \lVert \delta \psi' - \delta \psi \rVert^2= || \bar{O} \delta \theta - \tilde{\epsilon} ||^2,
\end{equation}
where $\tilde{\epsilon}=\Re \bar{\epsilon} + m i \Im \bar{\epsilon}$
is a modified vector and $m = \frac{\delta \tau_\phi}{\delta \tau_A} \geq 1$ is the step size ratio. Minimizing this modified distance, we obtain the following parameter update rules:
\begin{align}
    \text{SR: }\delta \theta &= (\bar{O}^\dagger \bar{O} + \beta \mathbb{I})^{-1} \bar{O}^\dagger \tilde{\epsilon} \label{eq:delta_theta_modified_SR}, \\
    \text{MinSR: }\delta \theta &= \bar{O}^\dagger (\bar{O}\bar{O}^\dagger + \beta \mathbb{I})^{-1} \tilde{\epsilon}. \label{eq:delta_theta_modified_MinSR}
\end{align}
This modification effectively slows the updates to the log-amplitude relative to the phase, which is crucial for learning unknown sign structure in challenging systems. Mathematically, Eq.~(\ref{eq:delta_theta_modified_SR}) and Eq.~(\ref{eq:delta_theta_modified_MinSR}) are equivalent. Both $(\bar{O}^\dagger \bar{O} + \beta \mathbb{I})$ and $(\bar{O}\bar{O}^\dagger + \beta \mathbb{I})$ are Hermitian positive definite matrices, which allow efficient inversion via Cholesky decomposition. However, these computations are prone to numerical instability in practice \cite{goldshlager2024kaczmarz}, necessitating double precision arithmetic for reliable computation. 

An alternative approach is to apply the scaling factor $m$ to the $\bar{O}$ matrix by defining $\tilde{O}=m\Re \bar{O} + i \Im \bar{O}$. While this leads to a similar formulation for SR $\delta \theta = (\tilde{O}^\dagger \tilde{O} + \beta \mathbb{I})^{-1} \tilde{O}^\dagger \bar{\epsilon}$, empirical results indicate that performance is improved by keeping the energy gradient $\bar{O}^\dagger \bar{\epsilon}$ unchanged. Consequently, we only use the modified matrix $\tilde{O}$ in the preconditioning term. This results in the update rules of preconditioned gradient:
\begin{align}
    &\text{SR: }\delta \theta = (\tilde{O}^\dagger \tilde{O} + \beta \mathbb{I})^{-1} \bar{O}^\dagger \bar{\epsilon} \label{eq:delta_theta_modified_SR_twotilde}, \\
    &\text{MinSR: }\delta \theta = \tilde{O}^\dagger (\tilde{O}\tilde{O}^\dagger +\beta \mathbb{I})^{-1} (\tilde{O}\tilde{O}^\dagger +\beta \mathbb{I})^{-1} \tilde{O}\bar{O}^\dagger \bar{\epsilon}.\label{eq:delta_theta_modified_MinSR_twotilde}
\end{align}

While this preconditioned approach lacks a physical interpretation as clear as the relative time-step approach derived above, it offers computational advantage as detailed in Section \ref{sec:discussion}. Further analysis for Eq.~(\ref{eq:delta_theta_modified_SR_twotilde}) and Eq.~(\ref{eq:delta_theta_modified_MinSR_twotilde}) are provided in the Supplementary Material. Finally, when the neural network parameters are real-valued but the outputs are complex, the $\bar{O}$ matrix and $\bar{\epsilon}$ vector need to be replaced by stacking their real and imaginary parts along the first axis \cite{chen2024empowering}.

\section{Discussion}
\label{sec:discussion}
We benchmark our methods on the spin-1/2 $J_1-J_2$ Heisenberg model with first- and second-neighbor couplings,

\begin{equation}
    H = J_1 \sum_{\langle i,j \rangle} \vec{s_i} \cdot \vec{s_j} 
      + J_2 \sum_{\langle \langle i,j \rangle \rangle} \vec{s_i} \cdot \vec{s_j}. 
\end{equation}
This model exhibits magnetic order at both small and large values of $J_2$, showing a N\'eel and a stripy pattern \cite{Sandvik_2010}, respectively. Near the classical maximally frustrated point, $J_2/J_1=0.5$, these orders vanish: current consensus is that this region is split into a $\mathbb{Z}_2$ spin liquid and a valence-bond solid (VBS) \cite{PhysRevB.108.054410,nomura2021dirac,liu2022gapless}. We estimate the ground-state energy on a $6 \times 6$ square lattice with periodic boundary conditions at $J_2/J_1=0$ (no frustration) and $J_2/J_1=0.5$ (maximal frustration).

A square lattice is a bipartite graph whose points can be divided into two disjoint and independent sublattices, A and B, such that every interaction bond connects a point in sublattice A to one in sublattice B. In the $J_2=0$ limit, where only nearest-neighbor interactions exist, one sublattice is a checkerboard pattern. In the $J_1=0$ limit, where only next-nearest-neighbor interactions exist, a sublattice has a striped pattern consisting of alternating rows or columns. Under these two limits, the sign of the wave function for a spin configuration $\sigma$ can be formally expressed as

\begin{equation}
    \text{sign}[\psi(\sigma)] = (-1)^{M_A(\sigma)},
\end{equation}
where $M_A$ represents the total number of up-spins in sublattice A for the configuration $\sigma$. This relation is known as the Marshall sign rule (MSR) \cite{marshall1955antiferromagnetism}. While this rule does not hold exactly for arbitrary values of $J_1$ and $J_2$, it serves as a good approximation for numerical calculations \cite{wang2024variational}.

\begin{figure*}[htbp]
    \centering
    \subfloat[]{\includegraphics[width=0.5 \textwidth]{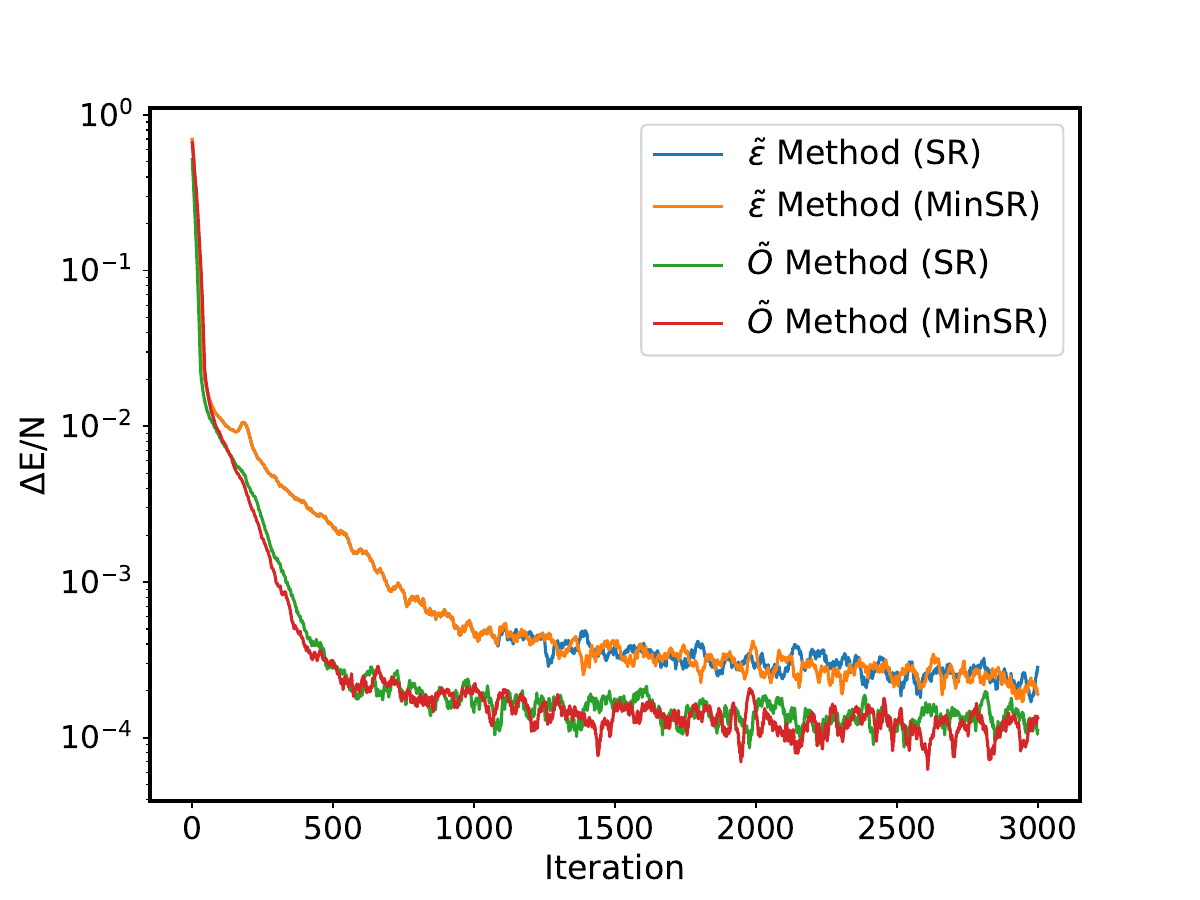}}
    \subfloat[]{\includegraphics[width=0.5 \textwidth]{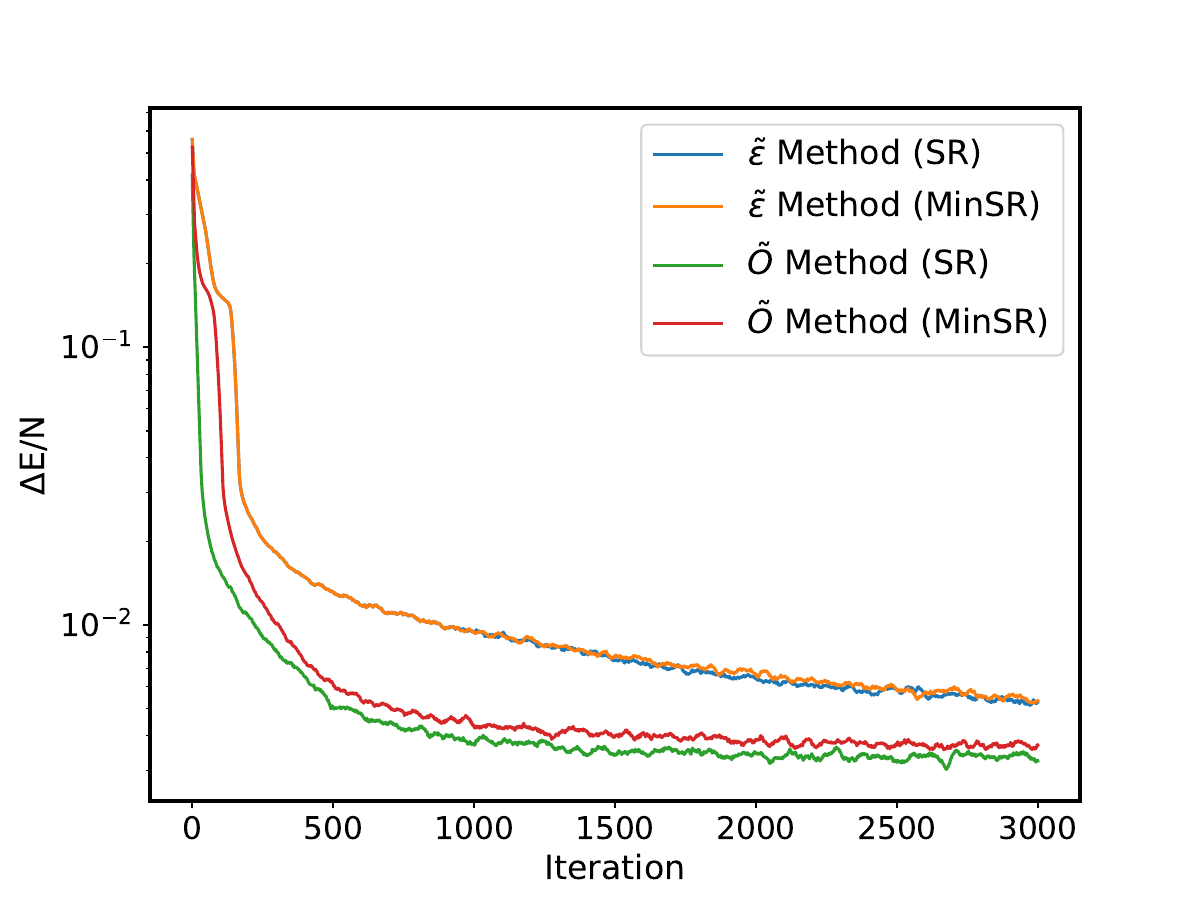}}
    \caption{Comparison of energy learning curves for (a) $J_2/J_1=0$ and (b) $J_2/J_1=0.5$. $\tilde{\epsilon}$ method (SR) refers to Eq.~(\ref{eq:delta_theta_modified_SR}) and $\tilde{\epsilon}$ method (MinSR) refers to Eq.~(\ref{eq:delta_theta_modified_MinSR}) with $m=4$, while $\tilde{O}$ method (SR) represents Eq.~(\ref{eq:delta_theta_modified_SR_twotilde}) and $\tilde{O}$ method (MinSR) represents Eq.~(\ref{eq:delta_theta_modified_MinSR_twotilde}) with $m=11$.}
    \label{fig:energy_pureSR_mass}
\end{figure*}

Figure \ref{fig:energy_pureSR_mass} compares the energy learning curves for $J_2/J_1=0$ and $J_2/J_1=0.5$ using three different approaches: the original SR method with $m=1$, our modified SR formulations (Eq.~(\ref{eq:delta_theta_modified_SR}) and Eq.~(\ref{eq:delta_theta_modified_SR_twotilde})), and our MinSR formulations (Eq.~(\ref{eq:delta_theta_modified_MinSR}) and Eq.~(\ref{eq:delta_theta_modified_MinSR_twotilde})). The choice of $m=1$ proves problematic, as it causes rapid collapse of Monte Carlo samples into two checkerboard patterns with the smallest diagonal Hamiltonian element. This collapse leads to zero gradient norm and traps the energy in a local minimum. In contrast, both our modified SR and MinSR formulations successfully achieve state-of-the-art accuracy for both coupling ratios. The non-frustrated scenario has higher accuracy than the frustrated case. Note that the $\tilde{\epsilon}$ method converges slower due to our use of a smaller fixed learning rate rather than an adaptive one, which we implement for numerical stability. With extended training, all methods are expected to converge to comparable accuracy levels. Details of experimental setup are provided in the Supplementary Material.

The scaling factor $m$ requires some tuning: a value too small impedes phase convergence, while a value too large can induce training instability. We find that $m = 4$ and $11$ are effective for the modified $\tilde{\epsilon}$ [Eqs.~(\ref{eq:delta_theta_modified_SR})--(\ref{eq:delta_theta_modified_MinSR})] and $\tilde{O}$ [Eqs.~(\ref{eq:delta_theta_modified_SR_twotilde})--(\ref{eq:delta_theta_modified_MinSR_twotilde})] methods, respectively.  The optimal $m$ is likely system- and learning-rate-dependent. Furthermore, it would be of interest to develop criteria for automatically changing $m$ during the simulation along the ideas proposed in \cite{goldshlager2024kaczmarz}.

Having established the effectiveness of our approach, we now present quantitative analyses using the best-performing method--Eq.~(\ref{eq:delta_theta_modified_SR_twotilde})--for both $J_2/J_1=0$ and $J_2/J_1=0.5$ cases. We compare ground state energies per site across different approaches. At $J_2/J_1=0$, our work yields $-0.6788$, showing excellent agreement with the exact value \cite{schulz1996magnetic} of $-0.678872$. At this coupling ratio, VMC method \cite{mezzacapo2009ground} gives $-0.6785(2)$, and CNN method \cite{choo2019two} achieves $-0.67882(1)$. For $J_2/J_1=0.5$, where the exact energy is $-0.503810$, our calculation produces $-0.5004$. The other methods produced the following results: VMC1 \cite{hu2013direct} obtained $-0.50117(1)$, VMC2 \cite{mezzacapo2009ground} reached $-0.4985(2)$, DMRG \cite{gong2014plaquette} achieved $-0.503805$, and CNN \cite{choo2019two} reported -0.50185(1).

\begin{figure}[htbp]
    \centering
    \includegraphics[width=0.45\textwidth]{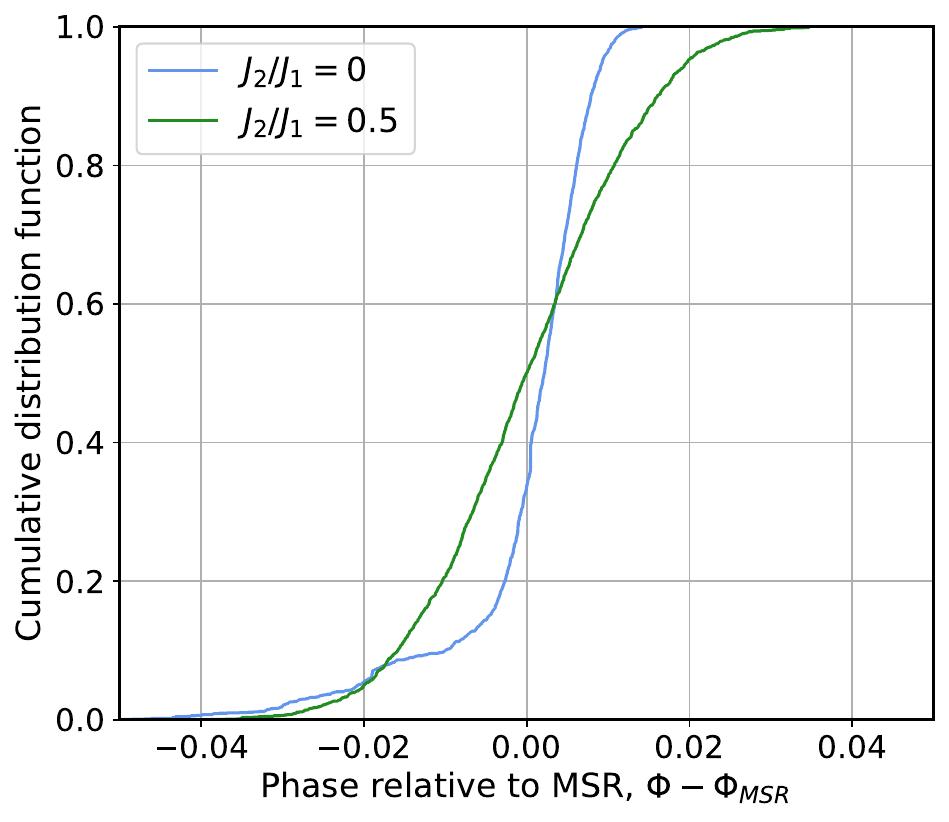}
    \caption{Cumulative distribution function of phase relative to MSR for $J_2/J_1=0$ and $J_2/J_1=0.5$.}
    \label{fig:phase_polar}
\end{figure}

Figure \ref{fig:phase_polar} shows the empirical cumulative distribution function for phase difference relative to MSR, $\phi-\phi_{MSR}$, for $J_2/J_1=0$ and $J_2/J_1=0.5$. The cumulative distribution function of the random variable $\phi$ evaluated at $x$ is the probability that $\phi-\phi_{MSR}$ takes a value less than or equal to $x$, i.e., $F(x)=\text{P}(\phi-\phi_{MSR} \leq x)$. Since two cases both have small standard deviations, our neural network with the improved stochastic reconfiguration method can effectively learn the Marshall sign rule.

For the frustrated case $J_2/J_1=0.5$, we calculate the overlap of the true sign structure and the Marshall sign to estimate the deviation,

\begin{equation}
    D = \Bigg | \sum_\sigma |\psi_0(\sigma)|^2 \text{sign}[\psi_0(\sigma)] \text{M}(\sigma) \Bigg |.
\end{equation}
where $\psi_0(\sigma)$ is the exact ground-state wave function from exact diagonalization, $\text{M}(\sigma)$ is the Marshall sign for configuration $\sigma$. The overall absolute value is used to take into account the overall phase difference. According to the definition, $0 \leq D \leq 1$. If $D=1$, then MSR is exactly true. If $D < 1$, the amount of difference from 1 indicates the total probability to have non-Marshall sign configurations. We find 49.76\% symmetrized basis configurations having non-Marshall sign, but get $D=0.9608$. This shows that non-Marshall sign configurations have small amplitudes and hence contribute little to the Monte Carlo sampled energy gradient. We hypothesize that the simple architecture of the phase network adopted in this study is not capable of representing the non-Marshall sign configurations, which do not follow any known simple rules that could be learned by a few layers of perceptrons.

In summary, we propose an improved stochastic reconfiguration method that can efficiently train the phase and amplitude concurrently by employing a faster learning rate for the phase. While this study concentrates on a spin model, the methods are straightforward to generalize to other fermionic and non-stoquastic Hamiltonians (a Hamiltonian is called stoquastic with respect to a basis if every off-diagonal element in that basis is non-positive; conversely, the ground state of a non-stoquastic Hamiltonian exhibits both positive and negative amplitudes, which poses a challenge in learning the correct sign structure). Furthermore, our approach can be extended to larger system sizes and more complex NQS architectures. The optimization scheme may also prove valuable in the context of variational Monte Carlo calculations that do not employ NQS wave functions. Finally, we point out that the symmetry-adapted method proposed in Appendix A can be useful in more complicated cases where the ground state (or the state of interest) lies at a nonzero total wave vector $\mathbf{k}$ and complex symmetry.



\begin{acknowledgments}
We gratefully acknowledge the use of computational resources at Yale Center for Research Computing. We thank Hongwei Chen and Douglas Hendry for drawing our attention to stochastic reconfiguration, and Ao Chen for leading us to the MinSR method. X.O. thanks Yuhang Li and Zejun Liu for helpful discussions.
\end{acknowledgments}

\appendix

\section{Symmetrized wave function}
\label{sec:symmetrized_wave_function}
There are two main approaches to implementing space group symmetry in neural network wave functions.  Let $G$ denote the space group with symmetry operations $\hat{g}$. The first step is to use data augmentation technique: we take each sampled state $\sigma$ and generate symmetry-related spin configurations by applying group operations, $\hat{g}\sigma$. The neural network is applied to all $\hat{g}\sigma$  separately, and the final result is taken as the average over this augmented symmetrized set. The same procedure is used for every initial state $\sigma$ which we seek to find the value of $\psi_\theta(\sigma)$. 

Alternatively, one can work with representative basis states. The idea is to pick a certain member $\sigma_E$ of the symmetry-equivalent set $\hat{g}\sigma$ ($\hat{g} \in G)$ as a representative and use it as the input to the neural network for all member of the set. This amounts to fixing the symmetry gauge by mapping each spin configuration in the Monte Carlo sample to its representative spin configuration, and in this way the network never sees spin configurations other than the representatives.

The first method is commonly believed to be superior than using representative spin configuration \cite{10.21468/SciPostPhys.10.6.147} because this exposes the neural net to a wider variety of configurations and imposes physically relevant constraints on the network parameters. In the main text, we choose the data augmentation method. However, our numerical experiment shows that these two methods indeed have similar performance in the energy minimization process. Furthermore, the representative ("canonical") configuration approach has advantages when the ground state is not of the completely symmetric $\boldsymbol{k}=0$ type and the proper way of averaging of neural network outputs over the set $\hat{g}\sigma$ is not self-evident.

Using representative configurations for implementing the full space group symmetry in the lattice is a general method for any $\boldsymbol{k}$ point and any irreducible representation. The starting point is the equivalence class of a basis state $\sigma$, usually referred to as the {\it orbit\/} of $\sigma$, given by $\mathcal{O}(\sigma)=\{\hat{g}\sigma, \forall \hat{g} \in G\}$. To distinguish between different orbits, we select a canonical configuration $\sigma_E$ for each. An orbit forms a $|\mathcal{O}|$-dimensional representation of $G$, which is in general reducible. The normalized symmetrized basis states of an orbit are calculated by
\begin{equation}
\label{normalized basis function}
    \ket{\phi_{mt}^j(\sigma_E)} = \sqrt{\frac{d_j}{|G|}} \sum_{g \in G} \Gamma^j(g)^*_{mt} \ket{\sigma_g},
\end{equation}
where we use $j \equiv (\boldsymbol{^*k},\beta)$ as a collective index for the k-points $\boldsymbol{k}$ and irrep $\beta$, $d$ is the dimension of the irrep, $\Gamma$ is the representation matrix, $m$ runs over all basis functions of the irrep, and $t$ represents the columns of $\Gamma$. We can now write the wave function with the normalized symmetrized basis states
\begin{equation}
    \ket{\Psi^j}=\sum_{\sigma_E,t} \Psi^j(\sigma_E,t) \ket{\phi^j_t(\sigma_E)},
\end{equation}
where the coefficients $\Psi^j(\sigma_E,t)$ represent the wave function in the symmetrized basis.

The final step is to establish the link between the wave function in the symmetrized basis, $\ket{\phi^j_t(\sigma_E)}$, and the neural network output $\Psi^{\text{net}}(\sigma_E)$. To obtain the coefficient $\Psi^j(\sigma_E,t)$, we apply the method of inversion in which we select a subset $G' \subset G$ such that the matrix
\begin{equation}\label{invertible matrix}
    M^j_{g't} = \sqrt{\frac{d_j}{|G|}}\Gamma^j(g')^*_{1t} \text{ with } g' \in G'
\end{equation}
is invertible, and use $M^j_{g't}$ to calculate the coefficient as
\begin{equation}\label{symmetrized wf}
    \mathrm{\Psi}^j(\sigma_E,t) = \sum_{g'\in G'} [M^j_{g't}]^{-1}\mathrm{\Psi}^\text{net}(\hat{g}'\sigma_E).
\end{equation}

As Eq.~(\ref{symmetrized wf}) suggests, we parameterize the symmetrized wave function by the canonical basis state $\sigma_E$, the k-points, and irreps. Further details of this approach are given in the Supplementary Material.

For the Heisenberg $J_1-J_2$ model with $J_2/J_1=0.5$, we compare the energy error per site using two different symmetrization approaches. The data augmentation method yields an energy error per site ($\Delta E/N$) of 0.0034, while the representative method produces an error of 0.0049. These two errors are of the same order of magnitude, indicating comparable performance between two symmetrization methods.

\bibliography{apssamp}

\end{document}


\maketitle

\section{Preconditioned energy gradient}
\subsection{Minimization of Eq.~(12)}
We start from the Lagrangian:
\begin{equation}
    \mathcal{L}_{\text{MinSR}}' = \beta \lVert \delta \theta \rVert^2 + (\delta \theta^\dagger \bar{O}^\dagger - \bar{\epsilon}^\dagger - \Delta^\dagger)\mu + \lVert \Delta \rVert^2.
    \label{eq:minsr-lagrangian}
\end{equation}

The optimality conditions are
\begin{align}
    &\frac{\partial \mathcal{L}_{\text{MinSR}}'}{\partial (\delta \theta^\dagger)} = \beta \delta \theta + \bar{O}^\dagger \mu = 0 \\
    &\frac{\partial \mathcal{L}_{\text{MinSR}}'}{\partial \Delta^\dagger} = -\mu+\Delta = 0 \\
    &\frac{\partial \mathcal{L}_{\text{MinSR}}'}{\partial \mu} = \delta \theta^\dagger \bar{O}^\dagger - \bar{\epsilon}^\dagger - \Delta^\dagger=0.
\end{align}
We have two equations after eliminating $\Delta=\mu$
\begin{align}
    &\beta \delta \theta + \bar{O}^\dagger \mu = 0 \label{eq:delta-theta}\\
    &\bar{O} \delta \theta = \bar{\epsilon} + \mu. \label{eq:mu}
\end{align}

These two equations have two solutions. For the first solution, we start by obtaining $\delta \theta = -\frac{1}{\beta} \bar{O}^\dagger \mu$ from Eq.~(\ref{eq:delta-theta}). Then we substitute this expression of $\delta \theta$ into Eq.~(\ref{eq:mu}), which gives $\mu = (-\beta) (\bar{O}\bar{O}^\dagger+\beta \mathbb{I})^{-1} \bar{\epsilon}$. Finally, replacing $\mu$ in Eq.~(\ref{eq:delta-theta}) yields the first solution 
\begin{equation}
    \delta \theta = \bar{O}^\dagger(\bar{O}\bar{O}^\dagger + \beta \mathbb{I})^{-1}\bar{\epsilon}. 
\end{equation}
This is the same as the MinSR solution except adding the Tikhonov regularization to stabilize the inversion.

Alternatively, one can first multiply $\bar{O}^\dagger$ from left on both sides of Eq.~(\ref{eq:mu}) and make use of Eq.~(\ref{eq:delta-theta}) to obtain $\bar{O}^\dagger \bar{O}\delta \theta = \bar{O}^\dagger \bar{\epsilon}+\bar{O}^\dagger \mu=\bar{O}^\dagger \bar{\epsilon}-\beta \delta \theta$. Solving this equation yields the second solution
\begin{equation}
    \delta \theta = (\bar{O}^\dagger \bar{O}+\beta \mathbb{I})^{-1}\bar{O}^\dagger \bar{\epsilon},
\end{equation}
which is the same as the SR solution with the Tikhonov regularization. The fact that both the SR and MinSR solutions can be derived from a single Lagrangian offers another way to see their mathematical equivalence.

\subsection{Derivation of Eq.~(19)}
From Eq.~(18), the change in the wave function with different imaginary time steps is

\begin{align}
    \delta \psi_\sigma &= \psi_\sigma (\frac{1}{A_\sigma}\frac{\partial A_\sigma}{\partial \tau} \delta \tau_A + i\frac{\partial \phi_\sigma}{\partial \tau} \delta \tau_\phi) \\
    &= \psi_\sigma (-\Re(E_{\text{loc}}(\sigma)-E)\delta \tau_A-i \Im(E_{\text{loc}}(\sigma)-E)  \delta\tau_\phi).
\end{align}
In contrast, the change due to a variation of the variational parameters is $\delta \psi'_\sigma =\psi_\sigma \sum_k (O_{\sigma k} - \langle O_{\sigma k} \rangle) \delta\theta_k$. The squared distance between these two wave function changes is

\begin{equation}
\begin{aligned}
    &\lVert \delta \psi' - \delta \psi \rVert^2 = \sum_\sigma |\psi_\sigma|^2 \Bigg[\\
    &\left( \sum_k (\Re O_{\sigma k} - \langle \Re O_{\sigma k} \rangle) \delta \theta_k + \Re(E_{\text{loc}}(\sigma)-E) \delta \tau_A  \right)^2 + \\
    &\left( \sum_k (\Im O_{\sigma k} - \langle \Im O_{\sigma k} \rangle) \delta \theta_k + \Im(E_{\text{loc}}(\sigma)-E) \delta \tau_\phi  \right)^2 \Bigg].
\end{aligned}
\end{equation}
By defining $\tilde{\epsilon}_\sigma=-\frac{\delta \tau}{\sqrt{N_s}}(\Re(E_{\text{loc}}(\sigma)-E) + m \Im(E_{\text{loc}}(\sigma)-E)) = \Re \bar{\epsilon}_\sigma+m i \Im \bar{\epsilon}_\sigma$, where $\delta \tau = \delta \tau_A$ and $m=\frac{\delta \tau_\phi}{\delta \tau_A}$, we obtain the following expression for the distance:

\begin{equation}
    d^2 = \lVert \delta \psi' - \delta \psi \rVert^2= || \bar{O} \delta \theta - \tilde{\epsilon} ||^2.
\end{equation}

Alternatively, we multiply the $m$ factor on the entire real term, which modifies the $\bar{O}$ matrix and maintains the definition of $\bar{\epsilon}_\sigma=-\delta \tau (E_{\text{loc}}(\sigma) - E )/\sqrt{N_s}$ intact.

\begin{equation}
\begin{aligned}
    &\lVert \delta \psi' - \delta \psi \rVert^2 \rightarrow \sum_\sigma |\psi_\sigma|^2 \Bigg[\\
    &\left( \sum_k (m\Re O_{\sigma k} - \langle m\Re O_{\sigma k} \rangle) \delta \theta_k + \Re(E_{\text{loc}}(\sigma)-E) \delta \tau  \right)^2 + \\
    &\left( \sum_k (\Im O_{\sigma k} - \langle \Im O_{\sigma k} \rangle) \delta \theta_k + \Im(E_{\text{loc}}(\sigma)-E) \delta \tau  \right)^2 \Bigg] \\
    &= || \tilde{O} \delta \theta - \bar{\epsilon} ||^2,
\end{aligned}
\end{equation}
where we introduce a modified matrix $\tilde{O}=m \text{Re}{\bar{O}}+i\text{Im}{\bar{O}}$ and set $\delta \tau_\phi \rightarrow \delta \tau$.

\subsection{Derivation of Eq.~(23)}
While using the modified logarithmic derivative of the wave function, $\tilde{O}$, we require the gradient to remain unchanged, i.e., $\tilde{O}^\dagger \bar{\epsilon}'=\bar{O}^\dagger \bar{\epsilon}$. However, this is a system of $N_p$ equations in $N_s$ variables $\bar{\epsilon}'$, hence overdetermined. We solve by the regularized Lagrangian:

\begin{equation}
    \mathcal{L}=\beta_1 \lVert \bar{\epsilon}' \rVert^2 + \lVert \tilde{O}^\dagger \bar{\epsilon}' - \bar{O}^\dagger \bar{\epsilon} \rVert^2,
\end{equation}
whose minimization yields 
\begin{equation}
\label{eq:epsilon-tilde}
\bar{\epsilon}'=(\tilde{O}\tilde{O}^\dagger +\beta_1 \mathbb{I})^{-1} \tilde{O}\bar{O}^\dagger \bar{\epsilon}.
\end{equation}

In the MinSR Lagrangian Eq.~(\ref{eq:minsr-lagrangian}), we now consider replacing $\bar{O}$ and $\bar{\epsilon}$ with their modified counterparts $\tilde{O}$ and $\bar{\epsilon}'$:
\begin{equation}
    \mathcal{L}_{\text{MinSR}}^\text{M} = \beta \lVert \delta \theta \rVert^2 + (\delta \theta^\dagger \tilde{O}^\dagger - \bar{\epsilon}'^\dagger - \Delta^\dagger)\mu + \lVert \Delta \rVert^2.
\end{equation}
Adopting the MinSR-type solution $\delta \theta = \tilde{O}^\dagger(\tilde{O}\tilde{O}^\dagger + \beta \mathbb{I})^{-1}\bar{\epsilon}'$ and substitute Eq.~(\ref{eq:epsilon-tilde}), we obtain
\begin{equation}
    \delta \theta = \tilde{O}^\dagger (\tilde{O}\tilde{O}^\dagger +\beta \mathbb{I})^{-1} (\tilde{O}\tilde{O}^\dagger +\beta_1 \mathbb{I})^{-1} \tilde{O}\bar{O}^\dagger \bar{\epsilon}.
\label{eq:delta_theta_modified_MinSR}
\end{equation}

Alternatively, using the SVD decomposition, we can demonstrate the mathematical equivalence of the two preconditioned matrices,
\begin{equation}
    (\tilde{O}^{\dagger}\tilde{O})^{-1} = \tilde{O}^{\dagger} (\tilde{O} \tilde{O}^{\dagger})^{-1} (\tilde{O} \tilde{O}^{\dagger})^{-1} \tilde{O}.
\end{equation}
The regularization parameters $\beta$ and $\beta_1$ are then added to ensure numerical stability. In practice, we set $\beta$ and $\beta_1$ to be equal.

\subsection{Nesterov's accelerated gradient method}

Nesterov's accelerated gradient (NAG), a seminal advancement in optimization algorithms, has emerged as a cornerstone in the realm of machine learning and numerical optimization. NAG can be viewed as the discrete version of a second-order ordinary differential equation (ODE) \cite{su2015differential},

\begin{equation}
    \frac{\dd^2 \theta}{\dd t^2} + \frac{3}{t} \frac{\dd \theta}{\dd t} + \nabla f (\theta)=0,
\end{equation}
where the time parameter $t$ relates to the step size. This formulation reveals an intuitive physical interpretation: the optimization dynamics resemble a damped system. At small $t$, the system is overdamped, smoothly converging to the equilibrium, while at larger $t$, the system becomes underdamped, exhibiting oscillations with gradually diminishing amplitude.

Given an objective function $f(\theta)$ to be minimized, the Nesterov's accelerated gradient update \cite{10.5555/3042817.3043064} is given by 

\begin{equation}
\begin{aligned}
    \vec{v}_{t+1} &= \mu \vec{v}_t -\epsilon \nabla f(\vec{\theta}_t+\mu \vec{v}_t), \\
    \vec{\theta}_{t+1} &= \vec{\theta}_t+\vec{v}_{t+1},
\label{eq:nesterov}
\end{aligned}
\end{equation}
where $\epsilon>0$ is the learning rate, $\mu \in [0,1]$ is the momentum coefficient. $\vec{\theta}_t$ and $\vec{v}_t$ are the parameter and momentum at optimization step $t$, respectively. A key difference between NAG and classical momentum (CM) lies in the gradient evaluation: NAG computes $\nabla f(\vec{\theta}_t + \mu \vec{v}_t)$, effectively "looking ahead" in the momentum direction, whereas CM uses $\nabla f(\vec{\theta}_t)$ from the current position $\theta_t$. This subtle modification enhances NAG's responsiveness, leading to more stable convergence, particularly for higher values of $\mu$ \cite{10.5555/3042817.3043064}. 

Beyond direct application, alternative approaches have been proposed to integrate momentum methods into stochastic reconfiguration framework \cite{goldshlager2024kaczmarz,gu2025solving}. These methods demonstrate the ability of momentum-based optimization in neural quantum states settings.

\subsection{Adaptive learning rate}
We also use an adaptive method to tune the learning rate, because convergence properties are strongly influenced by the choice of the learning rate: small values imply slow convergence rate, while large values may lead to oscillations or even divergence. The value of the learning rate is usually chosen by manual tuning and little attention has been paid to its automatic selection. Here, we propose to determine the learning rate for the subsequent step by maximizing the performance in the current step.

Suppose we have an objective function $f(\vec{\theta})$, where $\vec{\theta}$ is the set of parameters. With updated parameters, this function will be $f(\vec{\theta}+\tau \vec{\eta})$, where $\tau$ is learning rate and $\vec{\eta}$ is the search direction. We take first derivative to find the optimal learning rate $\tau_0$,

\begin{equation}
    \frac{\partial f}{\partial \tau} \Big|_{\tau=\tau_0} = \vec{\eta} \cdot \nabla f \Big|_{\vec{\theta}+\tau_0 \vec{\eta}} =0.
\end{equation}
In our current problem, the objective function $f(\vec{\theta})$ is energy $E(\vec{\theta})$, so that $\vec{\eta} \cdot \nabla E \Big|_{\vec{\theta}+\tau_0 \vec{\eta}} =0$. We define $\vec{\gamma} = \nabla E$ from now on.

After step $n$, the values of $\vec{\gamma}(\vec{\theta})$ and $\vec{\gamma} (\vec{\theta}+\tau^{(n)} \vec{\eta})$ are known, and we want to determine the optimal step size for the next iteration $n+1$, satisfying $\vec{\gamma}(\vec{\theta}+\tau^{(n+1)\vec{\eta}}) \cdot \vec{\eta}=0$. This can be accomplished by linear interpolation,

\begin{equation}
    \vec{\gamma}(\vec{\theta}) \cdot \vec{\eta} + \frac{\vec{\gamma} (\vec{\theta}+\tau^{(n)}\vec{\eta}) \cdot \vec{\eta} - \vec{\gamma} (\vec{\theta}) \cdot \vec{\eta}}{\tau^{(n)}} \tau^{(n+1)}=0.
\end{equation}

Therefore, the recursion formula can be written as

\begin{equation}
    \tau^{(n+1)} = \tau^{(n)} \frac{\vec{\gamma}(\vec{\theta}) \cdot \vec{\eta}}{\vec{\gamma}(\vec{\theta}) \cdot \vec{\eta} - \vec{\gamma} (\vec{\theta}+\tau^{(n)}\vec{\eta})\cdot \vec{\eta}}.
\end{equation}

In practice, $\vec{\gamma}(\vec{\theta}+\tau^{(n)} \vec{\eta})$ is estimated with updated wave function and local energies but with the old Monte Carlo sample. Notice that $\vec{\eta}$ is not recalculated with updated neural network parameters. If $0 < \tau^{(n+1)} < \tau^{(n)}$, the new learning rate is accepted; otherwise, we set $\tau^{(n+1)} = \min(1.2\tau^{(n)},0.2)$ to prevent the learning rate from growing too fast.

\begin{figure*}[htbp]
    \centering
    \includegraphics[width=0.6 \textwidth]{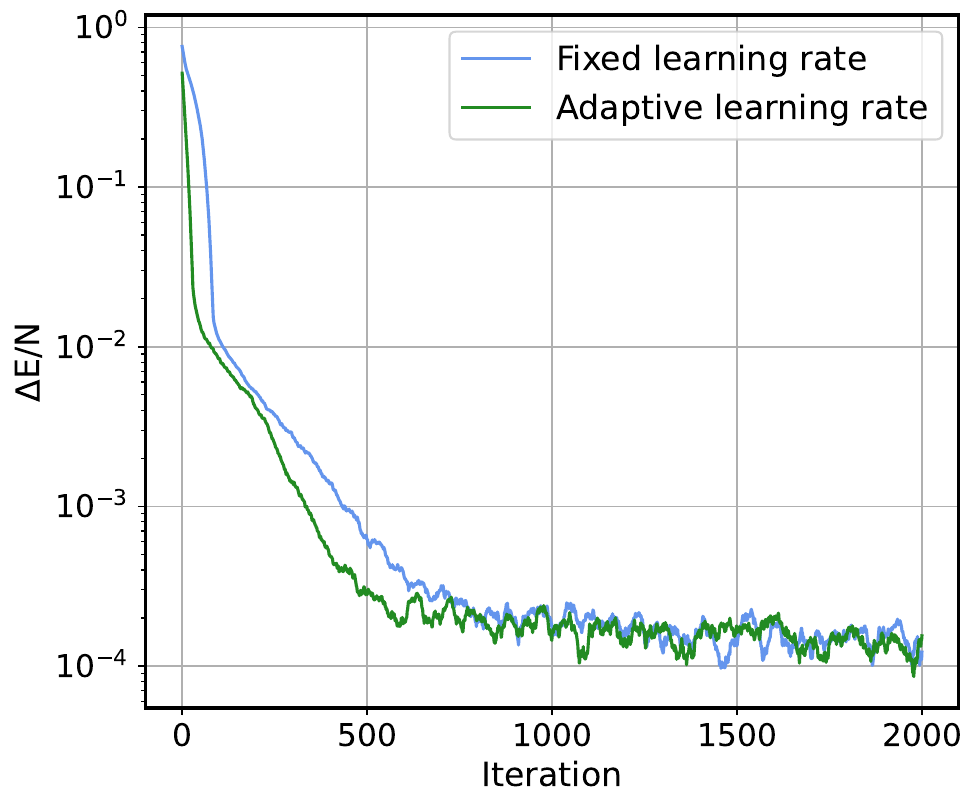}
    \caption{Comparison of energy learning curves with adaptive and fixed learning rate for $J_2/J_1=0$.}
    \label{fig:energy_adaptivestep}
\end{figure*}

Figure \ref{fig:energy_adaptivestep} compares energy learning curves with adaptive and fixed learning rate. Both Nesterov's accelerated gradient method and adaptive learning rate speed up the energy minimization process in the initial stage. With adaptive learning rate, the choice of momentum value $\mu$ doesn't make much difference.

\subsection{Hyperparameter}
Table \ref{tab:hyperparameter} and Table \ref{tab:hyperparameter-j205} summarize the experimental setups for the $J_2/J_1=0$ and $J_2/J_1=0.5$ cases, respectively. The methods referenced are defined as follows: the $\tilde{\epsilon}$ method by Eq.~(20) and Eq.~(21), the $\tilde{O}$ method (SR) by Eq.~(22), and the $\tilde{O}$ method (MinSR) by Eq.~(23) in the main text. These hyperparameter values were determined to be effective through numerical experiments.

\begin{table}[h]
\centering
\begin{tabular}{|c|c|c|c|}
\hline
\textbf{Config} & \textbf{$\tilde{\epsilon}$ Method} & \textbf{$\tilde{O}$ Method (SR)} & \textbf{$\tilde{O}$ Method (MinSR)} \\
\hline
Learning rate & 0.006 (fixed) & 0.04 (adaptive) & 0.04 (adaptive) \\
\hline
$m$ value & 11 & 4 & 4\\
\hline
$\beta$ value & 0.1 & 0.2 & 0.05 \\
\hline
\end{tabular}
\caption{Hyperparameter setup for the $J_2/J_1=0$ case.}
\label{tab:hyperparameter}
\end{table}

\begin{table}[h]
\centering
\begin{tabular}{|c|c|c|c|}
\hline
\textbf{Config} & \textbf{$\tilde{\epsilon}$ Method} & \textbf{$\tilde{O}$ Method (SR)} & \textbf{$\tilde{O}$ Method (MinSR)} \\
\hline
Learning rate & 0.009 (fixed) & 0.04 (adaptive) & 0.04 (adaptive) \\
\hline
$m$ value & 11 & 4 & 4 \\
\hline
$\beta$ value & 0.1 & 0.2 & 0.09 \\
\hline
\end{tabular}
\caption{Hyperparameter setup for the $J_2/J_1=0.5$ case.}
\label{tab:hyperparameter-j205}
\end{table}

\section{Space group symmetry}
Here, we outline our approach to incorporate the full space group symmetry in the lattice using representative configuration. We first introduce the basic background of space group, and then describe how the symmetrized basis functions are obtained.

\subsection{Background}

Consider a space group $\bf{G}$ with a subgroup $\bm{\mathrm{T}}=\sum_n \{E|\bm{t}_n\}$ of lattice translations written in the left coset decomposition,

\begin{equation}
    \bm{\mathrm{G}} = \{R_1|\bm{\tau}_1\}\bm{\mathrm{T}} + \{R_2|\bm{\tau}_2\}\bm{\mathrm{T}} + ... + \{R_h|\bm{\tau}_h\}\bm{\mathrm{T}},
\label{eq:left-coset-definition}
\end{equation}
where $R_j$ are point operations and $\bm{\tau}_j$ are fractional translations of least possible magnitude associated with $R_j$. In non-symmorphic space groups, the left coset representatives $\{R_j|\bm{\tau}_j\}$ themselves do not generally form a group, but the factor group $\bf{G}/\bf{T}$ is isomorphic with the point group $\bf{F}$ formed by $h$ elements $R_j$. The latter is called the isogonal point group of $\bf{G}$ and plays a key role in determining the symmetries of Bloch wave vectors and hence in classifying space group representations.

In fermionic systems, a basis function $\textbf{x}$ can be expressed as a Slater determinant:

\begin{equation}
  \textbf{x}_L(r_1,r_2,...,r_N;s_1,s_2,...,s_N) =\frac{1}{\sqrt{N!}}
  \begin{vmatrix}
    \phi_{l_1}^{\sigma_1}(r_1,s_1) & \phi_{l_1}^{\sigma_1}(r_2,s_2) & ... & \phi_{l_1}^{\sigma_1}(r_N,s_N) \\ 
    \phi_{l_2}^{\sigma_2}(r_1,s_1) & \phi_{l_2}^{\sigma_2}(r_2,s_2) & ... & \phi_{l_2}^{\sigma_2}(r_N,s_N) \\ 
    ... & ... & ... & ... \\ 
    \phi_{l_N}^{\sigma_N}(r_1,s_1) & \phi_{l_N}^{\sigma_N}(r_2,s_2) & ... & \phi_{l_N}^{\sigma_N}(r_N,s_N)
  \end{vmatrix},
\label{eq:basis-function-x}
\end{equation}

where $L=\{l_1,l_2,...,l_N;\sigma_1,\sigma_2,...,\sigma_N\}$ denotes the orbit label (site $l$ and spin $\sigma$), $\phi_{l_i}^{\sigma_i}(r_j) = \phi^\sigma_i(r_j-l_i)$ is the spatial orbital, $N$ is the total number of electrons, ($r_i$, $s_i$) represents the electron coordinate and spin. This formulation naturally extends to spin systems, which form a subset of fermionic models with a constraint that each site is occupied by one electron. In this case, the allowed moves must involve an even number of fermionic permutations (e.g., two-electrons exchanges). Consequently, the permutation signs are always $+1$, effectively resulting in Bose statistics.

A space group operation $\hat{g} = \{R|\bm{\tau}+\bm{t}\}$ acts on $\textbf{x}$ as
\begin{equation}
  \hat{g}\textbf{x}_L =\frac{1}{\sqrt{N!}}
  \begin{vmatrix}
    g\phi_{l_1}^{\sigma_1}(r_1,s_1) & g\phi_{l_1}^{\sigma_1}(r_2,s_2) & ... & g\phi_{l_1}^{\sigma_1}(r_N,s_N) \\ 
    g\phi_{l_2}^{\sigma_2}(r_1,s_1) & g\phi_{l_2}^{\sigma_2}(r_2,s_2) & ... & g\phi_{l_2}^{\sigma_2}(r_N,s_N) \\ 
    ... & ... & ... & ... \\ 
    g\phi_{l_N}^{\sigma_N}(r_1,s_1) & g\phi_{l_N}^{\sigma_N}(r_2,s_2) & ... & g\phi_{l_N}^{\sigma_N}(r_N,s_N)
  \end{vmatrix} = (-1)^{P_g} \textbf{x}'.
\end{equation}
$\hat{g}$ transforms $\textbf{x}$ into another basis function $\textbf{x}'$ by permuting the rows of the determinant, and $P_g$ denotes the number of permutations performed. For consistency, we must define a 'canonical order' that specifies how each row should be permuted. For example, in Equation (\ref{eq:basis-function-x}) the rows are permuted along $l_i$; for the same $l_i$, spin $\downarrow$ comes before spin $\uparrow$. The canonical order can be chosen in various ways so long as the ordering is consistent throughout the calculations.

We adopt an "active" convention for symmetry transformations. The transformation of the basis states works as follows:
\begin{equation}
    \hat{g}\ket{\textbf{x}} = \ket{\hat{g}\textbf{x}},
\end{equation}
and the operator $\hat{g}$ transforms an arbitrary many-body state $\ket{\mathrm{\Psi}} = \sum_x \mathrm{\Psi}(\textbf{x})\ket{\textbf{x}}$ as
\begin{equation}
    \hat{g}\ket{\mathrm{\Psi}} = \sum_x \mathrm{\Psi}(\textbf{x})\ket{\hat{g}\textbf{x}} = \sum_x \mathrm{\Psi}(\hat{g}^{-1}\textbf{x})\ket{\textbf{x}}.
\label{eq:transform-arbitratry-state}
\end{equation}
As Equation (\ref{eq:transform-arbitratry-state}) shows, to obtain $\hat{g}\ket{\mathrm{\Psi}}$ one can either transform the basis states $\ket{\textbf{x}}$ with fixed coefficients $\mathrm{\Psi}(\textbf{x})$, or transform the coefficient array.

The Bloch theorem for a single-particle state is written as
\begin{equation}
    \{E|\bm{t}\}\psi_{\bf{k}}(\bm{r}_0) = \psi_{\bf{k}}(\bm{r}_0-\bm{t}) = e^{-i\bf{k} \cdot \bm{t}}\psi_{\bf{k}}(\bm{r}_0),
\label{eq:Bloch-single-particle-state}
\end{equation}
and for the many-body state, we have
\begin{equation}
    \{E|\bm{t}\}\mathrm{\Psi}_{\bf{k}}(\textbf{x}) = e^{-i\bm{k} \cdot \bm{t}}\mathrm{\Psi}_{\bf{k}}(\textbf{x}).
\label{eq:Bloch-many-body-state}
\end{equation}
Equations (\ref{eq:Bloch-single-particle-state}) and (\ref{eq:Bloch-many-body-state}) show that the representations of $\bm{\mathrm{T}}$ are one-dimensional and labeled by the wave vector $\bf{k}$ from the first Brillouin zone (1BZ).

Applying a symmetry element $\{S|\bm{w}\}$ to a many-body Bloch function at wave vector $\bf{k}$ gives a Bloch function at wave vector $S\bf{k}$ (see Sec. 3.6 in \cite{bradley2009mathematical}):
\begin{equation}
    \{S|\bm{w}\}\mathrm{\Psi}_{\bf{k}}(\textbf{x}) = 
    \mathrm{\Psi}_{S\bf{k}}(\{S|\bm{w}\}^{-1}\textbf{x}).
\label{eq:transform-bloch-function}
\end{equation}

Now, suppose that $\bf{\Gamma}$ is a representation of $\bf{G}$; then the elements in $\bf{\Gamma}$ that represent translations form a representation of $\bf{T}$, which in general is reducible. Suppose that it reduces to representations characterized by wave vectors $\bf{k}_1,\bf{k}_2,...,\bf{k}_d$. As shown by Equation (\ref{eq:Bloch-many-body-state}), there exists a basis that consists of Bloch functions characterized by $\bf{k}_1,\bf{k}_2,...,\bf{k}_d$. Since these Bloch functions are eigenstates of $\bf{T}$, in this basis the elements of $\bf{T}$ are represented by diagonal matrices with diagonal elements given by $e^{-i\bf{k_pt}}$ (p = 1 to d). 

Equation (\ref{eq:transform-bloch-function}) implies we can obtain $\bf{k}_1,\bf{k}_2,...,\bf{k}_d$ by applying point group operations $R$ from the isogonal point group $\bf{F}$. For some vector $\bf{k}_1 \in $ 1BZ, we define the $\it{star}$ of $\bf{k}_1$ to be a set of distinct wave vectors obtained by applying point group operations:
\begin{equation}
    ^{*}\textbf{k} = {\bf{k}_1,\bf{k}_2,...,\bf{k}_q} = \bigcup_{R \in \bf{F}} R \bf{k_1}.
\end{equation}
Two wave vectors that differ by a reciprocal lattice vector $\bf{g}$ are considered equivalent, and we write $R\bf{k_1}\equiv \bf{k_1}$ if $R\bf{k_1} = \bf{k_1}+\bf{g}$.

The $\textit{little co-group}$ of $\bf{k_1}$ consists of those operations $S \in \bf{F}$ which leave $\bf{k_1}$ invariant or transforms it into an equivalent vector, i.e. $S \bf{k_1} \equiv \bf{k_1}$. We denote this subgroup of $\bf{F}$ by $\bar{\bf{G}}^{\bf_{k_1}}$. If $Q\bf{k_1} = \bf{k_2}$ is another member of the star then we have
\begin{equation}
    \bar{\bf{G}}^{\bf{k}_2} = Q\bar{\bf{G}}^{\bf{k}_1}Q^{-1},
\end{equation}
i.e., little co-groups at $\bf{k_1},\bf{k_2},...,\bf{k_q}$ form a set of conjugate subgroups of $\bf{F}$. Furthermore, the group $\bf{F}$ can be expressed as a sum of left cosets with respect to $\bar{\bf{G}}^{\bf{k}_1}$:
\begin{equation}
    \textbf{F} = E\bar{\bf{G}}^{\bf{k}_1} + Q_2 \bar{\bf{G}}^{\bf{k}_1} + ... + Q_q \bar{\bf{G}}^{\bf{k}_1},
\end{equation}
where $\bf{k_i}=Q_i \bf{k_1}$ and we have made the explicit choice $Q_1 = E$. These relations will be useful for constructing representations of the full space group $\bf{G}$.

Next we introduce the \textit{little group} $\bf{G}^{\bf{k}_1}$, which is a space group obtained as a subgroup of $\bf{G}$ in Equation (\ref{eq:left-coset-definition}) formed by left cosets with the rotational parts $E, S_2,...,S_b$ from the little co-group $\bar{\bf{G}}^{\bf{k}_1}$:
\begin{equation}
    \textbf{G}^{\bf{k}_1} = \{E|\bm{\tau}_1\}\textbf{T} + \{S_2|\bm{\tau}_2\}\textbf{T} + ... + \{S_b|\bm{\tau}_b\}\textbf{T}.
\label{eq:Gk1}
\end{equation}
Essentially, $\bf{G}^{\bf{k}_1}$ is the space group containing all symmetry operations of wave vector $\bf{k_1}$. We can then express the full space group as a sum of left cosets with respect to the little groups $\bf{G}^{\bf{k}_1}$:
\begin{equation}
    \textbf{G} = \{Q_1|\bm{\zeta}_1\}\textbf{G}^{\bf{k}_1} + \{Q_2|\bm{\zeta}_2\}\textbf{G}^{\bf{k}_1} + ... + \{Q_q|\bm{\zeta}_q\}\textbf{G}^{\bf{k}_1}.
\label{eq:left-coset-Gk1}
\end{equation}

A $\textit{small representation}$ $\mathbf{\Gamma}_\beta^{\bf{k_1}}$ of $\textbf{G}^{\bf{k}_1}$ is an irreducible representation of $\textbf{G}^{\bf{k}_1}$ whose basis functions satisfy the Bloch theorem in Equation (\ref{eq:Bloch-single-particle-state}) and (\ref{eq:Bloch-many-body-state}). Let the basis functions of $\mathbf{\Gamma}_\beta^{\bf{k_1}}$ with dimensionality $r$ be written as
\begin{equation}
    \mathrm{\Psi}_{\bf{k_1},1}^{\beta}(x), \mathrm{\Psi}_{\bf{k_1},2}^{\beta}(x),...,
    \mathrm{\Psi}_{\bf{k_1},r}^{\beta}(x).
\end{equation}
Then we can obtain an irreducible representation of the space group $\bf{G}$ by forming a linear closure of the vector space spanned by the functions
\begin{equation}
    \{Q_j|\bm{\zeta}_j\}\mathrm{\Psi}_{\bf{k_1},i}^{\beta}(x),
\end{equation}
where $j = \text{1 to } q$, $i = \text{1 to } r$, and $\{Q_j|\bm{\zeta}_j\}$ is any set of left coset representatives from Equation (\ref{eq:left-coset-Gk1}). The dimension of this representation is $d=qr=hr/b$.

This reduces the task of finding representations of $\bf{G}$ to the simpler task of finding small representations of $\bf{G}^{\bf{k}}$ for all $\bf{k}$ in the first Brillouin zone. The matrices of small representations of $\bf{G}^{\bf{k}}$ have the following simple form:
\begin{equation}
    \mathbf{\Gamma}_\beta^{\mathbf{k}} (\{R|\mathbf{v}\}) = e^{-i\mathbf{k\cdot v}} \mathbf{D}_\beta^{\bf{k}}(R),
\label{eq:gamma-definition}
\end{equation}
where $\mathbf{v} = \bm{\tau} + \bm{t}$ and $\mathbf{D}_\beta^{\bf{k}} (R)$ are the same for all left cosets in Equation (\ref{eq:Gk1}). $\mathbf{D}_\beta^{\bf{k}} (R)$ can be calculated with the method of projective representations \cite{bradley2009mathematical}, and they obey the following product rule:
\begin{equation}
    \mathbf{D}_\beta^{\bf{k}} (R_i)\mathbf{D}_\beta^{\bf{k}} (R_j) = e^{-i \bm{g}_i \cdot \bm{\tau}_j}\mathbf{D}_\beta^{\bf{k}} (R_i R_j),
\end{equation}
where $\bm{\tau}_j$ is the fractional translation associated with $R_j$ and $\bm{g}_i=R_i^{-1}\bf{k}-\bf{k}$.

\subsection{Symmetrized Supercell Basis}
In order to study physical models in reality, we construct supercells with lattice vectors that consist of primitive lattice vectors:
\begin{equation}
    \textbf{A}_i = \sum_j \textbf{M}_{ij} \textbf{a}_j,
\end{equation}
where $\textbf{A}_i$ denotes the supercell lattice vectors, $\textbf{M}_{ij}$ is a matrix of integers, and $\textbf{a}_j$ represents primitive lattice vectors. We use $\textbf{t}_n$ to denote the lattice vectors pointing to lattice sites $n = $ 1,...,$N_s$, where $N_S = |\textbf{M}|$ is the number of lattice sites. The translations $\textbf{t}_n$ form a finite group $\textbf{T}_S$ under periodic boundary conditions. Let $\textbf{G}_S$ denote the symmetry group of the supercell with symmetry operations $\hat{g_n} = \{R|\bm{\tau}+\textbf{t}_n\}$, which transforms the supercell sites into themselves.

Let $\textbf{x}$ denote the normalized unsymmetrized basis states, we define the orbit of $\textbf{x}$ to be
\begin{equation}
    \mathcal{O}(\textbf{x}) = \bigcup_{g_n \in \textbf{G}_S} \hat{g_n}\textbf{x}.
\end{equation}
$\mathcal{O}(\textbf{x})$ is the set of all distinct basis states that can be obtained by applying space group operations to $\textbf{x}$. The orbit is closed under $\textbf{G}_S$: any symmetry operation in $\textbf{G}_S$ would bring an element $\textbf{x} \in \mathcal{O}$ to another element $\textbf{x}' \in \mathcal{O}$. The set of all basis states can then be constructed by non-overlapping orbits. To distinguish between different orbits, we chose a "canonical" basis state $\textbf{x}_E$ in each orbit as a reference. $\textbf{x}_E$ can be defined arbitrarily as long as the definition is consistent and reproducible for all orbits. In practice, basis states are represented by arrays of integers. One way to define $\textbf{x}_E$ is to sort the basis states in the orbit and choose the nth state. We can now use $\textbf{x}_E$ to express other states in $\mathcal{O}(\textbf{x})$ as
\begin{equation}
    \textbf{x}_{g_n} = \hat{g}_n\textbf{x}_E.
\end{equation}

In this way, the basis states as well as their wave functions are related in an orbit, which reduces the number of unique wave function values by a factor of $|\mathcal{O}|$. The network now requires less trainable parameters to approximate the wave functions and thus improving the efficiency of network training.

\subsubsection{Symmetrized Wave Function}
An orbit forms a $|\mathcal{O}|$-dimensional representation of $\textbf{G}_S$ which is reducible in general. In order to utilize the symmetry properties of the wave functions, we express the normalized symmetrized basis functions of the orbit as
\begin{equation}
    \ket{\phi_{mt}^j(\textbf{x}_E)} = \sqrt{\frac{d_j}{|\textbf{G}_S|}} \sum_{g \in \textbf{G}_S} \mathbf{\Gamma}^j(g)^*_{mt} \ket{\textbf{x}_g},
\label{eq:normalized-symmetrized-basis-function}
\end{equation}
where we use $j \equiv (\textbf{*k},\beta)$ as a collective index for the k-points $\textbf{k}$ and irrep $\beta$, $d$ is the dimension of the irrep, $\mathbf{\Gamma}$ is the irrep matrix defined in Equation (\ref{eq:gamma-definition}), $m$ runs over all basis functions of the irrep, and $t$ represents the columns of $\mathbf{\Gamma}$. We can now write the wave function with the normalized symmetrized basis states:
\begin{equation}
    \ket{\mathrm{\Psi}^j} = \sum_{\textbf{x}_E,t} \mathrm{\Psi}^j(\textbf{x}_E,t) \ket{\phi^j_{t}(\textbf{x}_E)}.
\label{eq:any-state-in-symmetrized-basis}
\end{equation}

Substituting Equation (\ref{eq:normalized-symmetrized-basis-function}) into (\ref{eq:any-state-in-symmetrized-basis}), a general state can be expressed as
\begin{align}
\begin{split}
    \ket{\mathrm{\Psi}^j} &= \sum_{\textbf{x}_E,t} \mathrm{\Psi}^j(\textbf{x}_E,t) \sqrt{\frac{d_j}{|\textbf{G}_S|}} \sum_{g \in \textbf{G}_S} \mathbf{\Gamma}^j(g)^*_{1t} \ket{\textbf{x}_g}\\
    & = \sum_{\textbf{x}_E,g\in\textbf{G}_S} 
    \left( \sum_{t}\mathrm{\Psi}^j(\textbf{x}_E,t)\sqrt{\frac{d_j}{|\textbf{G}_S|}}\mathbf{\Gamma}^j(g)^*_{1t} \right)\ket{\textbf{x}_g}\\
    &= \sum_{\textbf{x}_E,g\in\textbf{G}_S} \mathrm{\Psi}^\text{net}(\textbf{x}_E,g)\ket{\textbf{x}_g},
\end{split}
\label{eq:symmetrized-wave-function}
\end{align}
where we select $m=1$, and expression in the parenthesis can be represented by the network representation of the wave function $\mathrm{\Psi}^\text{net}(\textbf{x}_g)$. The coefficient $\mathrm{\Psi}^j(\textbf{x}_E,t)$ can be obtained by method of "inversion", where we select a subset $\textbf{G}_S' \subset \textbf{G}_S$ such that the matrix
\begin{equation}
    \textbf{M}^j_{g't} = \sqrt{\frac{d_j}{|\textbf{G}_S|}}\mathbf{\Gamma}^j(g')^*_{1t} \text{ with } g' \in \textbf{G}_S'
\end{equation}
is invertible, and use $\textbf{M}^j_{g't}$ to calculate
\begin{equation}
    \mathrm{\Psi}^j(\textbf{x}_E,t) = \sum_{g'\in\textbf{G}_S'} [\textbf{M}^j]^{-1}_{g't}\mathrm{\Psi}^\text{net}(\hat{g}'\textbf{x}_E).
\end{equation}
Therefore, we can express any quantum state in the normalized symmetrized basis states with the symmetrized wave functions, which brings a great reduction in the dimension of the Hilbert space.

\subsubsection{Symmetrized Energy, Loss and Gradient}
Given the symmetrized wave function in Equation (\ref{eq:symmetrized-wave-function}), we can write the symmetrized average energy as
\begin{equation}
    E^j = \sum_{\{\textbf{x}_E,t\} \in \textbf{S}_{E\neq}} w^j (\textbf{x}_E,t) E^j(\textbf{x}_E,t),
\end{equation}
where $\textbf{S}_{E\neq}$ is the set of unique canonical basis states, $w^j(\textbf{x}_E,t)$ is the normalized weight which approximates the probability of the basis state $(\textbf{x}_E,t)$, and $E^j(\textbf{x}_E,t)$ denotes the local energy at state $(\textbf{x}_E,t)$. The normalized weights can be approximated by
\begin{equation}
    w^j(\textbf{x}_E,t) = \frac{|\mathrm{\Psi}^j(\textbf{x}_E,t)|^2}{\sum_{\{\textbf{x}_E',t'\} \in \textbf{S}_{E\neq}}|\mathrm{\Psi}^j(\textbf{x}_E',t')|^2}.
\end{equation}
The local energy is given by
\begin{equation}
    E^j(\textbf{x}_E,t) = \mathrm{\Psi}^j(\textbf{x}_E,t)^{-1}  \sum_{\textbf{y}_E} \sum_{s=1}^{n_{\textbf{y}_E}^{j}} H_{ts}^{j}(\textbf{x}_E,\textbf{y}_E) \mathrm{\Psi}^j(\textbf{y}_E,s),
\label{eq:local-energy}
\end{equation}
where the Hamiltonian $H_{ts}^j$ and states $\textbf{y}_E$ are described in detail in the next subsection.

The gradient of the average energy is modified to
\begin{equation}
    \partial_w E^j \approx 2 \sum_{\{\textbf{x}_E,t\} \in \textbf{S}_{E\neq}} w^j(\textbf{x}_E,t) \text{Re}[(E^j(\textbf{x}_E,t)-E^j)\mathrm{\Psi}^j(\textbf{x}_E,t)\partial_w \mathrm{\Psi}^j(\textbf{x}_E,t)^*],
\end{equation}
and the loss function is defined to be
\begin{equation}
    L^j = \sum_{\{\textbf{x}_E',t'\} \in \textbf{S}_{E\neq}} g^j(\textbf{x}_E,t) \mathrm{\Psi}^j(\textbf{x}_E,t)^*.
\end{equation}
where $g^j(\textbf{x}_E,t) = 2 w^j(\textbf{x}_E,t) \text{Re}[(E^j(\textbf{x}_E,t)-E^j)\mathrm{\Psi}^j(\textbf{x}_E,t)]$.

The main difference is that now the network manages to find the lowest energy state with respect to a specific k-point and irrep. 

\subsubsection{Symmetrized Hamiltonian}
To calculate the local energy in Equation (\ref{eq:local-energy}), we need to obtain the Hamiltonian matrix $H_{ts}^j$ and states $\textbf{y}_E$ for a given canonical states $\textbf{x}_E$. Let a general Hamiltonian matrix element be expressed as
\begin{align}
\begin{split}
    \bra{\phi_{mt}^j(x_E)}\hat{H}\ket{\phi_{ns}^i(y_E)} &= 
    \sqrt{\frac{d_i d_j}{|\textbf{G}_S|^2}} \sum_{g,h} \mathbf{\Gamma}^j(g)_{mt}\mathbf{\Gamma}^i(h)^*_{ns} 
    \mel{\hat{g}x_E}{\hat{H}}{\hat{h}y_E} \\
    &= \sqrt{\frac{d_i d_j}{|\textbf{G}_S|^2}} \sum_{g,h} \mathbf{\Gamma}^j(g)_{mt}\mathbf{\Gamma}^i(gh)^*_{ns} 
    \mel{x_E}{\hat{H}}{\hat{h}y_E}.
\end{split}
\end{align}
The second line is obtained by rewriting the matrix element as $\mel{\hat{g}x_E}{\hat{H}}{\hat{h}y_E}=\mel{x_E}{\hat{g}^{-1}\hat{H}\hat{g}}{\hat{g}^{-1}\hat{h}y_E}=\mel{x_E}{\hat{H}}{\hat{g}^{-1}\hat{h}y_E}$, and substituting $\hat{g}^{-1}\hat{h} \Rightarrow \hat{h}$.

We can expand $\mathbf{\Gamma}^i(gh)^*$ into a matrix product of $\mathbf{\Gamma}^i(g)^*$ and $\mathbf{\Gamma}^i(h)^*$ to get
\begin{align}
\begin{split}
    \bra{\phi_{mt}^j(x_E)}\hat{H}\ket{\phi_{ns}^i(y_E)} &= 
    \sqrt{\frac{d_i d_j}{|\textbf{G}_S|^2}} \sum_{p}\sum_{g,h} \mathbf{\Gamma}^j(g)_{mt} \mathbf{\Gamma}^i(g)^*_{np} \mathbf{\Gamma}^i(h)^*_{ps}
    \mel{x_E}{\hat{H}}{\hat{h}y_E} \\
    &= \delta^{ij}\delta_{mn} \sum_{h} \mathbf{\Gamma}^i(h)^*_{ts}\mel{x_E}{\hat{H}}{\hat{h}y_E} \\
    &= \sqrt{\frac{|\textbf{G}_S|}{d_i}} \delta^{ij}\delta_{mn} \mel{x_E}{\hat{H}}{\phi_{ts}^j(y_E)}.
\end{split}
\end{align}

The result indicates that the Hamiltonian becomes block-diagonal with blocks indexed by $j \equiv (\textbf{k}^*,\beta)$ and irrep row indices $m$. Each block has matrix elements indexed by orbit states $(x_E,y_E)$ and irrep basis sets $(s,t)$. The steps to calculate the symmetrized Hamiltonian are described in Algorithm \ref{alg:symmetrized-hamiltonian}.

\begin{algorithm}
\linespread{1.35}\selectfont
\caption{Symmetrized Hamiltonian. $x_E$ is the canonical state of which we calculate the Hamiltonian, $t$ is the corresponding column of the irrep matrix, $y$ represents the state that has a nonzero Hamiltonian element with $x_E$.}
Inputs: 
$x_E, t$\;
1. Find all $y$ where $\mel{x_E}{\hat{H}}{y}\neq 0$\;
2. Find $y_E$ for all $y$ and set $H_{ts} (x_E,y_E)=0$\;
3. Calculate the normalization factors for $n = x_E, y_E$ by
$N_n = \sqrt{\sum_{g\in\textbf{G}_S}\mathbf{\Gamma}^j(g)^*_{tt}} $\;
\For{each $y$}{
a). Find all $\hat{h}$ such that $y=\hat{h}y_E$, then $\mel{x_E}{\hat{H}}{y} = \mel{x_E}{\hat{H}}{\hat{h}y_E}$\;
b). Update matrix elements by \\
$H^j_{ts}(\textbf{x}_E,\textbf{y}_E) \mathrel{+}= \frac{1}{N_{x_E},N_{y_E}} \left(\sqrt{\frac{d_j}{|\textbf{G}_S|}}\sum_{h\in\{\hat{h}y_E=y\}}\mathbf{\Gamma}^j(h)^*_{ts}\right)\mel{x_E}{\hat{H}}{y}$\;
}
Outputs: All $y_E$ and $H_{ts}^j$\;
\label{alg:symmetrized-hamiltonian}
\end{algorithm}

\bibliographystyle{plain}
\bibliography{main}